\documentclass[conference,twocolumn,romanappendices]{IEEEtran}
\def\paperversion{2} 

\IEEEoverridecommandlockouts

\usepackage[subtle,indent=normal,mathspacing=normal]{savetrees}

\usepackage{balance}
\usepackage[normalem]{ulem}

\usepackage{cite}
\usepackage{algorithmic}
\usepackage{graphicx}
\usepackage{textcomp}
\usepackage{xcolor}

\usepackage[cmex10]{amsmath}
\usepackage{amssymb}
\usepackage{amsthm}
\usepackage{amsfonts}
\usepackage{mathtools}
\usepackage{bm}
\usepackage{stmaryrd}
\usepackage{url}
\usepackage{enumerate}
\usepackage{booktabs}
\usepackage{multirow}
\usepackage{colortbl}
\usepackage[font=footnotesize]{caption}
\usepackage[font=footnotesize]{subcaption}

\usepackage[T1]{fontenc}
\usepackage{todonotes}
\usepackage{float}

\usepackage{tikz}
\usepackage{pgfplots}
\pgfplotsset{my style/.append style={axis x line=middle, axis y line=
middle, xlabel={$x$}, ylabel={$y$}, axis equal }}

\newtheorem{theorem}{Theorem}
\newtheorem{corollary}{Corollary}
\newtheorem{proposition}{Proposition}
\theoremstyle{definition}
\theoremstyle{remark}
\theoremstyle{remark}

\newcommand{\trans}{^{\mathsf T}}
\newcommand{\hath}{\widehat{\mathfrak{h}}}
\newcommand{\h}{\mathfrak{h}}
\newcommand{\llb}{\llbracket}
\newcommand{\rrb}{\rrbracket}
\DeclarePairedDelimiterX{\inp}[2]{\langle}{\rangle}{#1, #2}
\DeclareMathOperator*{\argmin}{\arg\!\min}

\title{Small-Sample~Inferred Adaptive Recoding for Batched Network Coding}
\author{\IEEEauthorblockN{Jie Wang, Zhiyuan Jia, Hoover H. F. Yin, and Shenghao Yang$^\dagger$}
\thanks{J.~Wang, Z.~Jia and S.~Yang are with the School of Science and Engineering, The Chinese University of Hong Kong, Shenzhen, Shenzhen, China.
H.~Yin is with the Institute of Network Coding, The Chinese University of Hong Kong, Hong Kong, China.
J.~Wang is also with the H. Milton Stewart School of Industrial and Systems Engineering, Georgia Institute of Technology, Atlanta, USA.
S.~Yang is also with the Shenzhen Key Laboratory of IoT Intelligent Systems and Wireless Network Technology, Shenzhen, China. This work was funded in part by the Shenzhen Science and Technology Innovation Committee (Grant ZDSYS20170725140921348, JCYJ20180508162604311).
}
\thanks{$^\dagger$Corresponding author. Email: shyang@cuhk.edu.cn}
}

\newcommand{\N}{\llbracket N \rrbracket}
\begin{document}

\maketitle

\begin{abstract}
Batched network coding is a low-complexity network coding solution to feedbackless multi-hop wireless packet network transmission with packet loss.
	The data to be transmitted is encoded into batches where each of which consists of a few coded packets.
	Unlike the traditional forwarding strategy, the intermediate network nodes have to perform recoding, which generates recoded packets by network coding operations restricted within the same batch.
	Adaptive recoding is a technique to adapt the fluctuation of packet loss by optimizing the number of recoded packets per batch to enhance the throughput.
	The input rank distribution, which is a piece of information regarding the batches arriving at the node, is required to apply adaptive recoding.
	However, this distribution is not known in advance in practice as the incoming link's channel condition may change from time to time.
	On the other hand, to fully utilize the potential of adaptive recoding, we need to have a good estimation of this distribution.
	In other words, we need to guess this distribution from a few samples so that we can apply adaptive recoding as soon as possible.
	In this paper, we propose a distributionally robust optimization for adaptive recoding with a small-sample inferred prediction of the input rank distribution.
	We develop an algorithm to efficiently solve this optimization with the support of theoretical guarantees that our optimization's performance would constitute as a confidence lower bound of the optimal throughput with high probability.
\end{abstract}

\section{Introduction}
	 
	For some communication applications in extreme environments such as deep space \cite{breidenthal2000merits} and underwater \cite{sozer2000underwater}, enabling feedback can be expensive so that feedbackless multi-hop wireless packet networks are cost-efficient in practice.
	Due to different reasons, including interference and signal fading, packet loss is a common phenomenon in wireless networks.
	Although \emph{fountain codes} \cite{lubyLT,shokRaptor,maymounkov02} can recover the lost packets without the need for feedbacks by their ratelessness property, their rates are low in such multi-hop networks because a packet can be received by the destination node if it is not lost at any of the lossy links.
	On the other hand, \emph{random linear network coding (RLNC)} \cite{random,random2,jaggi03,Sanders03}, a simple realization of \emph{linear network coding} \cite{flow,alg,linear}, is well-known to have throughput gain over forwarding in general without depending on feedback or knowledge of the network topology.
	The issue on huge computational and storage costs at the intermediate network nodes can be resolved by using a variation of RLNC called \emph{batched network coding} \cite{chou03,Silva2009,Heidarzadeh2010,Mahdaviani12,yang14bats}.
	There exist batched network codes that can achieve close-to-optimal rates, e.g., BATS codes \cite{yang14bats,bats_book}, which are suitable to be deployed in the aforementioned extreme environments, i.e., in deep space \cite{zhao16ds,bittt_space} and in underwater \cite{underwater,sprea2019bats}.
	 
	A batched network code encodes the data to be transmitted into \emph{batches} where each of which consists of a few coded packets, with each packet attached with a coefficient vector.
	The number of linearly independent coefficient vectors in a batch, which is the amount of information carried by the batch, is called the \emph{rank} of the batch.
	As a type of network codes, the intermediate network nodes have to perform re-encoding, or simply called \emph{recoding}, instead of forwarding.
	The recoding of a batched network code is restricted within the same batch, which generates recoded packets by performing RLNC on the received packets of the batch.
	The simplest recoding scheme, called \emph{baseline recoding}, is to generate the same number of recoded packets regardless of the ranks of the batches, but the throughput is not optimal in general \cite{yang14a}.
	 
	\emph{Adaptive recoding} \cite{scheduling,uni,rf,ge_adaptive,adaptive} is an advanced recoding scheme that optimizes the throughput at the next node by assigning different number of recoded packets for batches of different ranks.
	In other words, the optimization requires knowledge of the ranks of the received batches.
	The distribution of these ranks is called the \emph{input rank distribution}.
	Unless the channel conditions of all links keep unchanged over time and we know the conditions exactly, we cannot precompute the input rank distributions at each of the network nodes.
	On the other hand, we need to have at least an estimation of the input rank distribution in order to calculate the number of recoded packets by adaptive recoding.
	That is, we need to predict an input rank distribution from a few received batches to reduce the delay induced before we can decide the number of recoded packets.
	One simple approach is to group a few batches into a block and perform adaptive recoding block by block \cite{adaptive,yin21impact,yin21intrablock}.
	However, \cite{adaptive} also showed that a larger block size results in better throughput, i.e., the input rank distribution should capture the ranks of the batches out of the observation in order to maximize the throughput.
	 
	In this work, we consider a scenario where the observation is short so that the empirical input rank distribution may not be a faithful representation of the underlying distribution.
	The implementation of adaptive recoding based on this empirical distribution would lead to disappointment in out-of-sample throughput performance.
	The Bayesian inference is a sample-efficient approach to estimate the input rank distribution, but a wrong choice of the prior information can result in non-negligible biases for distributional estimation, which significantly impacts the throughput.
	Instead, we follow the frequentist's principle by considering the \emph{distributionally robust optimization (DRO)} framework \cite{gao2016distributionally}, which modifies the original adaptive recoding problem in a way that optimizes the number of recoded packets under the most adverse input rank distributions within an ambiguity set.
	By calibrating the ambiguity sets carefully, the optimized number of recoded packets for DRO can give high out-of-sample throughput.
	We develop an algorithm to efficiently solve this joint optimization with the support of theoretical guarantees that the optimal value of DRO provides a lower confidence bound on the achievable out-of-sample throughput.
\ifnum\paperversion=1
Omitted proofs can be found in \cite{wang2021smallsample}.
\else
The proofs of this paper can be found in the Appendix.
\fi


\textit{Notations:} 
Denote by $\mathbb{E}$ the expectation operator.
For any non-negative integer $N$, define $[N] := \{0, 1, \ldots, N\}$.
For any positive integer $N$, define $\llb N \rrb := \{1,2,\ldots,N\}$.
Fix a positive integer $M$.
Define $\delta_{x} := (\delta_{x,0},\delta_{x,1},\ldots,\delta_{x,M})$ as a $M$-vector of Kronecker deltas.
For a function $f \colon [M] \to \mathbb{R}$ and a probability distribution $\mathfrak{h}$ supported on $[M]$, denote by
\[\textstyle
\|f\|_{\text{Lip}, \mathfrak{h}}
:=
\max\left\{
\frac{|f(\tilde{r}) - f(r)|}{|\tilde{r} - r|} \colon 
r\in\text{supp}(\mathfrak{h}), \tilde{r}\in[M], \tilde{r}\ne r
\right\}
%
\]
the Lipschitz norm of $f$ with respect to $\mathfrak{h}$.
Denote the $1$-Wasserstein distance by 
\[\textstyle
W(\mathfrak{h}_1, \mathfrak{h}_2)=\min_{\gamma}~\int_{[M]\times[M]} |r_1-r_2|\gamma(dr_1, dr_2),
\] 
where $\gamma$ is a joint distribution on $[M]\times[M]$ with marginal distributions $\mathfrak{h}_1$ and $\mathfrak{h}_2$. 
Denote by $\bm0_N$ and $\bm1_N$ the column zero vector and column all-ones vector of length $N$ respectively.
For two vectors $\mathbf{a}$ and $\mathbf{b}$ of the same length, their maximum $\max(\mathbf{a}, \mathbf{b})$ is taken component-wisely.


\section{Adaptive Recoding}

Suppose we want to send a file from a source node to a destination node through multiple intermediate nodes in a feedbackless packet network with packet loss.
For the sake of reliability of data transmission, we adopt batched network codes in this network.

\subsection{Batched Network Coding}

We divide the file to be sent into multiple \emph{input packets} where each of which has the same length and is regarded as a vector over a fixed finite field.
At the source node, we apply the encoder of a batched network code which generates $M$ coded packets per \emph{batch} from the input packets, where $M$ is a small positive integer called the \emph{batch size}. 
	The batch size is not necessarily a constant for all batches \cite{tree,variable}, although most works assume a constant batch size for simplicity.
The selection of input packets to constitute the coded packets depends on the batched network code.
Each coded packet is formed by taking a random linear combination of the selected input packets.
A coefficient vector is attached to each coded packet for recording the linear network coding operations at the intermediate network nodes.
Two packets in the same batch are linearly independent of each other if and only if their coefficient vectors are linearly independent of each other.
The \emph{rank} of a batch is the number of linearly independent coefficient vectors in it.
By manipulating the coefficient vectors, we make a freshly generated batch to have rank $M$.

When the batches travel through the network, their ranks are monotonically decreasing due to packet loss.
At each intermediate network node, instead of forwarding, we perform \emph{recoding}, which generates recoded packets for each batch by taking random linear combinations of the received packets of this batch.
By generating extra recoded packets which act as redundancy, we can reduce the rank-losing rate, where the rank can be interpreted as the information carried by the batch.

At the destination node, we apply the decoder of the batched network code to rewind the linear operations applied on the received packets so that the input packets can be recovered.
Except Gaussian elimination, belief propagation decoding and inactivation decoding \cite{Raptormono,inactivation} are alternative decoding algorithms if the batched network code supports them.

\subsection{Adaptive Recoding}

We cannot transmit recoded packets for a single batch indefinitely, so we have to decide how many recoded packets to be generated with a constraint that the average number of recoded packets among all the batches is $t_\text{avg}$ for some $t_\text{avg}$ which can maintain a stable queue of packets at the intermediate network node.
The simplest \emph{baseline recoding} generates the same number of recoded packets for all batches regardless of the ranks of the batches.
However, we can see intuitively that this scheme is not optimal in throughput \cite{yang14a}, because a batch of higher rank is likely to lose some rank when the redundancy is not enough, while a batch of lower rank is likely to preserve its rank when the redundancy is too large.
In other words, we should carefully assign the number of recoded packets for each batch.
\emph{Adaptive recoding} is an approach for this purpose which only depends on the local knowledge so that it can be applied distributively at the intermediate network nodes.

Let $E_r(t)$ be the expected rank of a batch at the next node when we transmit $t$ recoded packets of this batch, and this batch has rank $r$ at the current node.
For simplicity, we call the expected rank at the next node the \emph{expected rank}.
The exact formulation of the expected rank, which can be found in \cite{uni}, depends on the condition of the channel towards the next node. 
As there is no feedback enabled, we cannot know whether there is a change in the channel condition, thus we assume that the channel condition is stationary.
This way, we know from \cite{uni} that $E_r(t)$ is concave, which can be interpreted as, at the next node, the chance of a newly received packet being linearly independent of the already received packets decreases when the (expected) number of received packets increases.
Further, we have $0 \le E_r(t) \le r$ and $E_r(t)$ is monotonically increasing.

To simplify the notation, we follow \cite{uni} to endow the meaning of non-integer $t$: we transmit $\lfloor t \rfloor + 1$ packets with probability $t - \lfloor t \rfloor$ and transmits $\lfloor t \rfloor$ packets with probability $1- (t - \lfloor t \rfloor)$.
We have
\begin{equation*}
	E_r(t) = (t-\lfloor t \rfloor) E_r(\lfloor t \rfloor + 1) + (1-t+\lfloor t \rfloor) E_r(\lfloor t \rfloor).
\end{equation*}

Fix an intermediate network node.
Let $M$ be the maximum rank among all the batches.
Denote by $\mathfrak{h}:=\{h_r\}_{r\in[M]}$ the distribution of the ranks of the batches arriving at the node,
which is also called the \emph{input rank distribution}.
Adaptive recoding obtains the number of recoded packets $t_r$ for a batch of rank $r$ by solving
\begin{equation}
\tag{IP} \label{eq:IP}
\max_{\bm t} \mathbb{E}_{r\sim\mathfrak{h}}~[E_r(t_r)]\quad
\mathrm{s.t.} \quad \mathbb{E}_{r\sim\mathfrak{h}}~[t_r]=t_\text{avg},
\end{equation}
where $\bm t:=\{t_r\}_{r\in[M]}$ is the vector of numbers of recoded packets for different ranks $r\in[M]$, called the \emph{recoding vector}.
\subsection{Issues about Input Rank Distribution}

It is impractical to obtain the exact information on the communication channel.
For instance, a burst loss channel cannot be perfectly modeled by a multiple-state Markov chain \cite{nstate_fail}.
In this case, we may use an estimation model instead,
 e.g., Gilbert-Elliott model \cite{GilbertBurst,ElliottBurst} for bursty channel, or independent packet loss model to imitate a burst loss channel with an interleaver.
Consequently, the lack of knowledge on the channel condition forbids us to calculate the exact input rank distribution analytically.

Although we can obtain an accurate empirical input rank distribution by receiving many batches, we cannot wait until we have such an accurate distribution because we have to solve \eqref{eq:IP} as soon as possible to minimize the delay induced.
On the other hand, an input rank distribution restricting on a few newest batches cannot result in the best throughput \cite{adaptive}.
That is, the input rank distribution for \eqref{eq:IP} should capture the ranks of the batches out of the observation, including the batches to be received in the future.
In other words, we have to predict this distribution from a few received batches, which leads to our discussion in the rest of this paper.

\section{Distributionally Robust Optimization}

Denote by $\hath_N:=\frac{1}{N}\sum_{j=1}^N\delta_{\hat{r}_j}$ the empirical rank distribution based on $N$ collected samples $\hat{r}_i, i\in \llb N \rrb$, where
the empirical estimate $\hath_N$ of $\h$ actually achieves the lower bound up to constant factors for the minimax risk~\cite{Singh2020EstimatingPD}:
\[
\mathcal{R}_N=\inf_{\hath}\sup_{\h\in\mathcal{P}}~
\mathbb{E}_{\hat{r}_i\sim \h}~[W(\mathfrak{h}, \hat{\mathfrak{h}}(\hat{r}_1,\ldots,\hat{r}_N))],
\]
where $\mathcal{P}$ denotes the set of discrete probability distributions with the support $[M]$, and the infimum is taken over all possible estimators $\hath \colon [M]^N \to \mathcal{P}$.
Therefore, regardless of \eqref{eq:IP}, the empirical distribution $\widehat{\mathfrak{h}}_N$ is the optimal choice for estimating the underlying rank distribution.
Provided that the underlying rank distribution is unknown, it is natural to approximately solve \eqref{eq:IP} by replacing $\mathfrak{h}$ with its empirical distribution $\widehat{\mathfrak{h}}_N$, called the \emph{sample average approximation (SAA)} method, while it suffers from the disappointment of out-of-sample performance.

We consider the distributionally robust adaptive recoding formulation, where the goal is to find a recoding vector that maximizes the worst-case utility function in a way that the worst-case expected number of recoded packets should not exceed the given resource $t_{\text{avg}}$. More specifically, given a recoding vector $\bm t=\{t_r\}_{r\in[M]}$, define the worst-case utility function as
\[\textstyle
\mathcal{U}(\bm t\mid\mathcal{P}_1):=
\inf_{\mathfrak{h}\in\mathcal{P}_1}~\mathbb{E}_{r\sim \mathfrak{h}}\left[E_r(t_r)\right].
\]
Also, define the worst-case expected number of recoded packets as
\[\textstyle
\mathcal{E}(\bm t\mid\mathcal{P}_2):=
\sup_{\mathfrak{h}\in\mathcal{P}_2}~\mathbb{E}_{r\sim \mathfrak{h}}[t_r].
\]
Based on the ambiguity sets $\mathcal{P}_1$ and $\mathcal{P}_2$, 
we would like to find the optimal solution to the following problem
\begin{equation}
\tag{DRO-IP}\label{eq:DRO:IP}
\sup_{\bm t}~
\mathcal{U}(\bm t\mid\mathcal{P}_1)
\quad \mathrm{s.t.}
\quad
\mathcal{E}(\bm t\mid\mathcal{P}_2)\le t_{\text{avg}},
\end{equation}
where $\mathcal{P}_1$ and $\mathcal{P}_2$ are ambiguity sets containing a collection of distributions around the empirical distribution $\widehat{\mathfrak{h}}_N$.
Since any distribution $\mathfrak{h}\in\mathcal{P}_i$ for $i = 1, 2$ should be close to the empirical distribution $\hath_N$ in the sense of some proper statistical distance, we adopt the Wasserstein distance to construct the ambiguity sets as
\[\textstyle
\mathcal{P}_i
=\{
\mathfrak{h}:~
W(\mathfrak{h}, \hath_N)\le\rho_i
\},
\]
where $\rho_i$ is determined such that the underlying input rank distribution is contained in the ambiguity set. 
Therefore, the ambiguity set $\mathcal{P}_i$ contains all probability distributions whose Wasserstein distances to the empirical input rank distribution $\hath_N$ are no more than $\rho_i$. 
The Wasserstein distance has applications in a variety of areas such as hypothesis testing~\cite{Mueller15, wang2021twosample, wang2021kerneltwosample} and statistical learning~\cite{Martin17, tolstikhin2018wasserstein, gao2020wasserstein}.
In our formulation, this distance naturally considers the geometry of the adaptive recoding space $[M]$. Moreover, it is well-defined even if two distributions have non-overlapping supports.
Finally, the ambiguity sets are purely data-driven because the nominal distribution $\hath_N$ is constructed based on collected samples, as opposed to moment ambiguity sets discussed in \cite{Ye10, Goh10, Wolfram14}.
This kind of Wasserstein distributionally robust framework generalizes better when applied out of sample.
In the following, we develop a tractable formulation for \eqref{eq:DRO:IP} and design a customized algorithm to solve it efficiently.
Finally, we discuss the choice of the radius sizes $\rho_1$ and $\rho_2$ with performance guarantees.

Before proceeding, we provide a different interpretation for \eqref{eq:DRO:IP} from the regularization perspective.
The proof is by reformulating $\mathcal{U}(\bm t\mid\mathcal{P}_1)$ and $\mathcal{E}(\bm t\mid\mathcal{P}_2)$ using the recent result \cite[Proposition~6]{wang2021reliable} on the equivalence between Wasserstein DRO and Lipschitz norm regularization.
\begin{proposition}\label{Proposition:regularization}
For fixed $\bm t$, denote by $E^{\bm t} \colon [M]\to\mathbb{R}$ a mapping with $r\mapsto E_r(t_r), r\in[M]$.
There exists $\bar{\rho}_i > 0$ such that for all $\rho_i<\bar{\rho}_i, i=1,2$, the problem~\eqref{eq:DRO:IP} is equivalent to 
\[
\begin{IEEEeqnarraybox}[][c]{rCl}
		\max_{\bm t\ge0} & \quad & \mathbb{E}_{r\sim\hath_N}~[E_r(t_r)] - \rho_1\|E^{\bm t}\|_{\text{Lip}, \hath_N} \\
		\mathrm{s.t.} && \mathbb{E}_{r\sim\hath_N}~[t_r] + \rho_2\|\bm t\|_{\text{Lip}, \hath_N}
		 \le t_\text{avg}.
	\end{IEEEeqnarraybox}
	\]
\end{proposition}

Besides constructing ambiguity sets using the Wasserstein
distance, the equivalence between regularization and DRO with other types of distances has also been studied in the literature.
For example, DRO problem with $\phi$-divergence ambiguity sets is asymptotically equivalent to variance regularization \cite{GOTOH2018448, lam2015robust, duchi2018statistics}.
On the contrary, Proposition 1 reveals an exact equivalence between \eqref{eq:DRO:IP} and its Lipschitz norm regularization.

\subsection{Tractable Formulation}
The current formulation in \eqref{eq:DRO:IP} is intractable since the evaluations on $\mathcal{U}(\bm t\mid\mathcal{P}_1)$ and $\mathcal{E}(\bm t\mid\mathcal{P}_2)$ take account of infinite numbers of possible distributions. By utilizing the duality result in \cite{gao2016distributionally}, we first reformulate it as a finite-dimensional optimization problem.
\begin{theorem}\label{Theorem:DRP:reformulate:1}
The max-min problem in \eqref{eq:DRO:IP} is equivalent to the following problem:
\begin{subequations}
\begin{align*}
\sup_{
\substack{
\bm t\ge0,\\
\lambda_{0,1}, \lambda_{0,2}\ge0
}}&\quad -\lambda_{0,1}\rho_1 + \frac{1}{N}\sum_{j=1}^N\inf_{r\in[M]}\bigg(
E_r(t_r) + \lambda_{0,1} |r-\hat{r}_j|
\bigg)\\
\mbox{s.t.}&\quad \lambda_{0,2}\rho_2 + \frac{1}{N}\sum_{j=1}^N\sup_{r\in[M]}\bigg(
t_r - \lambda_{0,2} |r - \hat{r}_j|
\bigg)\le t_{\text{avg}}.
\end{align*}
\end{subequations}
\end{theorem}

Theorem~\ref{Theorem:DRP:reformulate:1} holds regardless of the form of the expected rank function $E_r(\cdot)$.
As shown in \cite[Theorem~1]{uni}, $E_r(\cdot)$ is a monotonically increasing concave function.
The throughput increases when $t_r$ increases. However, a large value of $t_r$ spends many resources but does not increase the objective function too much, which limits the performance of adaptive recoding~\cite{scheduling}.
Therefore, we assume that the optimal number of recoded packets satisfies $0\le t_r\le i_{\max}^r$ for any $r\in[M]$, where $i_{\max}^r$ is an integer. 
Then we are able to represent the expected rank function $E_r(t)$ as a piecewise linear function.
\begin{theorem}\label{Theorem:E_r:piecewise:linear}
The expected rank function has the equivalent formulation $E_r(t)=\min_{i\in[i_{\max}^r]}~(\Delta_{r,i}t + \zeta_{r,i})$,
where 
$t \in [0,i_{\max}^r]$,
$\Delta_{r,i} := E_r(i+1)-E_r(i)$, and
$\zeta_{r,i} := E_r(i) - i\Delta_{r,i}$.
\end{theorem}
By substituting the piecewise linear expression of the expected rank function into the optimization problem in Theorem~\ref{Theorem:DRP:reformulate:1} and introducing slack variables, we obtain an equivalent linear programming~(LP) formulation. 
\begin{corollary}\label{Corollary:DRO:LP:reformula}
The problem~\eqref{eq:DRO:IP} admits an equivalent linear programming formulation:
\begin{equation}\label{eq:DRO:LP:reformulate}
\begin{aligned}
\max_{\substack{
\bm t\ge0\\
\lambda_{0,i}\ge0, \lambda_{j,i}, j\in\llb N\rrb, i=1,2
}} 
\!\!\!\!\!\!\!\!\!\!\!\!\!\!\! 
&\quad -\lambda_{0,1}\rho_1 + \frac{1}{N}\sum_{j=1}^N\lambda_{j,1}\\
\mbox{s.t.}&\quad \lambda_{0,2}\rho_2 + \frac{1}{N}\sum_{j=1}^N\lambda_{j,2}\le t_{\text{avg}},\\
&\quad \lambda_{j,1}\le \Delta_{r,i}t_r + \zeta_{r,i} + \lambda_{0,1} |r - \hat{r}_j|,
\nonumber\\
&\qquad\qquad
\ j\in\llb N\rrb, i\in[i_{\max}^r], r\in[M],\\
&\quad \lambda_{j,2}\ge t_r - \lambda_{0,2} |r - \hat{r}_j|,
\\
&\qquad\qquad
\ j\in\llb N\rrb, r\in[M]. \nonumber
\end{aligned}
\end{equation}
\end{corollary}

\subsection{Customized Algorithm}
Although the LP formulation 
can be solved by off-the-shelf solvers such as CVX \cite{cvx}, 
we develop a customized algorithm to solve this problem with lower computational complexity.
We define the following notations to express the LP formulation in Corollary~\ref{Corollary:DRO:LP:reformula} as the compact matrix form.
Define the primal variable $x=(t\trans,\lambda_1\trans,\lambda_2\trans)\trans$, where
\begin{align*}
t&=(t_0,\ldots,t_{M})\in\mathbb{R}^{M+1},\\
\lambda_i&=(\lambda_{0,i},\ldots,\lambda_{N,i})\trans\in\mathbb{R}^{N+1},\ i=1,2.
\end{align*}
Define the coefficient vector 
$f=(\bm0_{M+1}\trans,\rho_1,-\frac1N\bm1_{N}\trans, \bm0_{N+1}\trans)\trans$.
Then, define the constraint matrix and constraint vector as follows:

\noindent\scalebox{0.93}{\parbox{\linewidth}{%
\begin{align*}
\Delta_{r}&=(\Delta_{r,0},\ldots,\Delta_{r,i_{\max}^r})\trans, 
\zeta_{r}=(\zeta_{r,0},\ldots,\zeta_{r,i_{\max}^r})\trans, \forall r\in[M]\\
\overline{\Delta}&=\text{diag}(\Delta_{r}), \overline{\zeta}=(\zeta_{0}\trans,\ldots,\zeta_{M}\trans)\trans,
R=\text{vec}\big( |r-\hat{r}_j| \big)_{r\in[M], j\in\N} \\
R_j&=\text{vec}\big( |r-\hat{r}_j| \big)_{r\in[M]}, E_j=e_j\trans\otimes\bm1_{(M+1)(I+1)}, \forall j\in\llb N\rrb\\
A^{(0)}&=
(\bm0_{M+N+2}, \rho_2, \frac1N\bm1_{N})\\
A^{(1)}_j&=(-\overline{\Delta}, -R_j\otimes1_{I+1}, E_j, \bm0_{(M+1)(I+1)\times(N+1)}),\ \forall j\in\llb N\rrb
\\
A^{(2)}&=
(
\bm 1_N\otimes\bm I_{M+1}, \bm0_{N(M+1)\times(N+1)}, -R, -\bm I_{N}\otimes\bm1_{M+1}
)\\
A_{\text{ineq}}&=((A^{(0)})\trans, (A^{(1)}_1)\trans, \ldots, (A^{(1)}_N)\trans, (A^{(2)})\trans)\trans\\
b_{\text{ineq}}&=(t_{\text{avg}}, \overline{\zeta}\trans, \bm0_{N(M+1)}\trans)\trans.
\end{align*}
}}

As a result, the formulation in Corollary~\ref{Corollary:DRO:LP:reformula} can be expressed as
\begin{equation}
\tag{DRO-LP}\label{eq:DRO:LP}
x^*=\argmin\limits_{x\in\Gamma}~
f\trans x
\quad \mathrm{s.t.}
\quad
A_{\text{ineq}}x\le b_{\text{ineq}},
\end{equation}
where $\Gamma=\{x\in\mathbb{R}^{M+2N+3} \colon x_{1:M+2}\ge0, x_{M+N+3}\ge0\}$.
Considering that the constraint matrix $A_{\text{ineq}}$ is ill-conditioned, we apply the preconditioning technique to \eqref{eq:DRO:LP} to improve the numerical convergence behavior of our designed algorithms.
In particular, we solve the problem
\begin{equation}
\tag{DRO-LP-Scaled}\label{eq:DRO:LP:Scaled}
y^*=\argmin\limits_{y\in\Gamma}~
(f')\trans y
\quad \mathrm{s.t.}
\quad
A_{\text{ineq}}'y\le b_{\text{ineq}}',
\end{equation}
where $f'=fD_{\mathcal{R}}$, $A_{\text{ineq}}'=D_{\mathcal{L}}A_{\text{ineq}}D_{\mathcal{R}}$, and $b_{\text{ineq}}'=D_{\mathcal{L}}b_{\text{ineq}}$.
We design the diagonal matrices $D_{\mathcal{L}}, D_{\mathcal{R}}$ with positive diagonal entries by using the \emph{arithmetic mean} criteria~\cite{Ploskas17} to decrease the variance between the nonzero elements in the constraint matrix $A_{\text{ineq}}$.
After the optimal solution $y^*$ for \eqref{eq:DRO:LP:Scaled} is obtained, the optimal solution for \eqref{eq:DRO:LP} is recovered as $x^*=y^*D_{\mathcal{R}}$.

Now we discuss the design of a customized algorithm for solving the LP problem \eqref{eq:DRO:LP:Scaled} efficiently. 
The Lagrangian formulation of this problem is a convex-concave saddle point problem:
\[
\min_{y\in \Gamma}\max_{z\ge0}~\ell(y) + z\trans Ay - g(z),
\]
where $\ell(y)=(f')\trans y$, $g(z)=(b')\trans z$, and $A=A_{\text{ineq}}'$.
We use the \emph{adaptive primal-dual hybrid gradient method}~(aPDHG) to solve the saddle problem presented above, in which the iteration updates are the following~\cite{Goldstein2015}:
\begin{subequations}
\begin{align}
\hat{y}^{k+1}&=y^k - \tau_k A\trans z^k,\label{eq:update:matrix:1}\\
y^{k+1}&=\textstyle\argmin_{y\in\Gamma}~
\left\{
\ell(y) + \frac{1}{2\tau_k}\|y - \hat{y}^{k+1}\|^2
\right\},\label{eq:update:y}\\
\hat{z}^{k+1}&=z^k+\sigma_k A(2y^{k+1}-y^k),\label{eq:update:matrix:2}\\
z^{k+1}&=\textstyle\argmin_{z\ge0}~\left\{
g(z) + \frac{1}{2\sigma_k}\|z - \hat{z}^{k+1}\|^2
\right\}\label{eq:update:z}
\end{align}
in which the updates \eqref{eq:update:y} and \eqref{eq:update:z} have closed form solutions:
\[
y^{k+1}=\max(\hat{y}^{k+1}-\tau_k f',\text{lb}),\quad
z^{k+1}=\max(\hat{z}^{k+1}-\sigma_k b',0),
\]
with $\text{lb}=(\bm0_{M+2}\trans, -\infty\bm1_{N}\trans,0,-\infty\bm1_{N}\trans)\trans$. 
\end{subequations}
The step sizes $\{\tau_k\}$ and $\{\sigma_k\}$ in aPDHG are updated using the backtracing rule together with the residual balancing technique.

The major advantage of aPDHG over the existing algorithms can be summarized as follows.
First, the iteration updates involved in the aPDHG have closed-form solutions.
Second, by utilizing the block structure of the matrix $A$, the matrix multiplication operations involved in \eqref{eq:update:matrix:1} and \eqref{eq:update:matrix:2} can be implemented efficiently.
Moreover, many algorithms require a careful choice of the step-size parameters. On the contrary, the self-adaptive step size rules in aPDHG can automatically tune the hyper-parameters for optimal convergence.
Optimization convergence theory in \cite[Theorem~1]{Goldstein2015} guarantees the convergence of our designed algorithm with a rate $\mathcal{O}(1/k)$, where $k$ denotes the iteration number.

\subsection{Performance Guarantees}
\begin{figure*}
\includegraphics[width=0.23\textwidth]{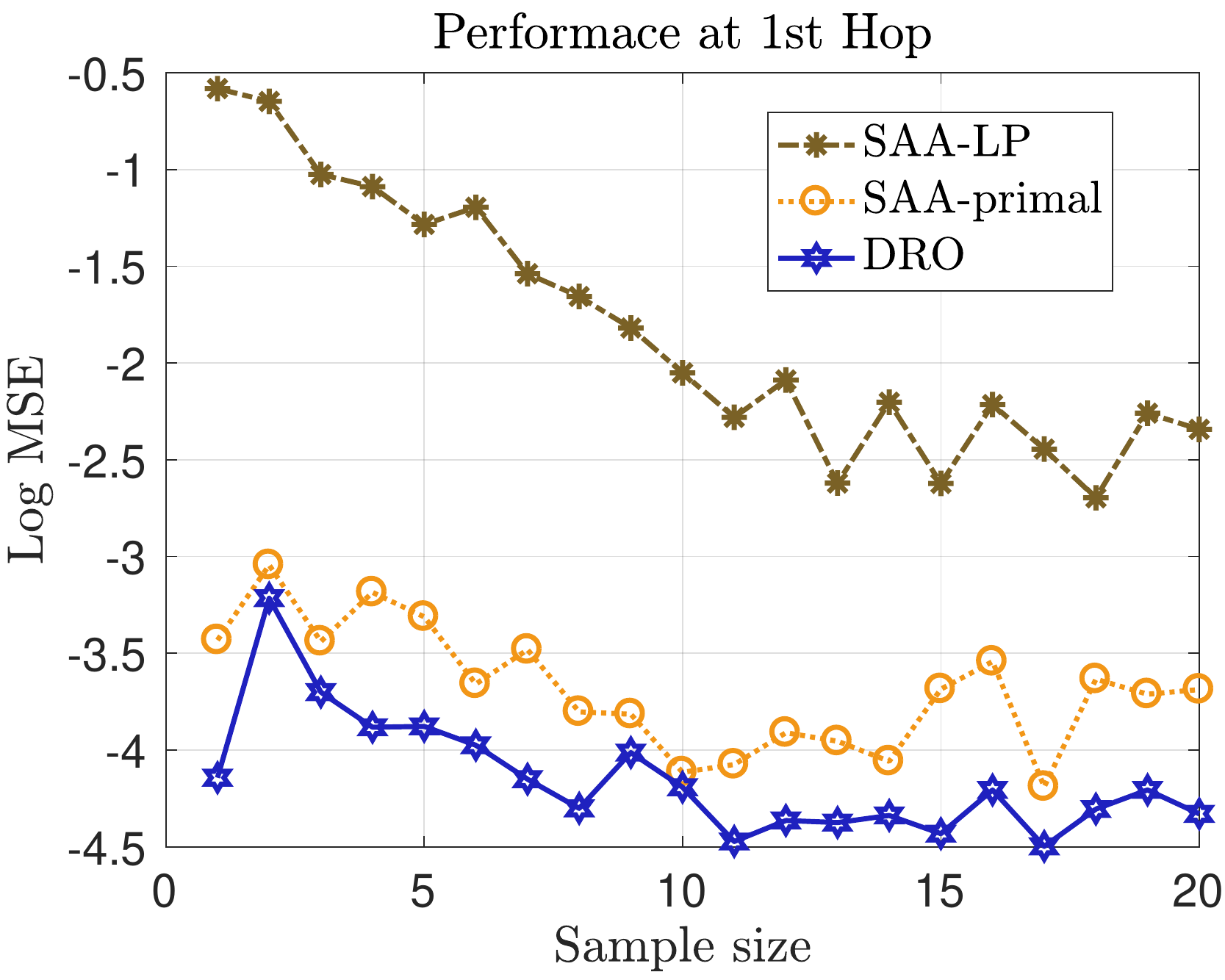}
\includegraphics[width=0.23\textwidth]{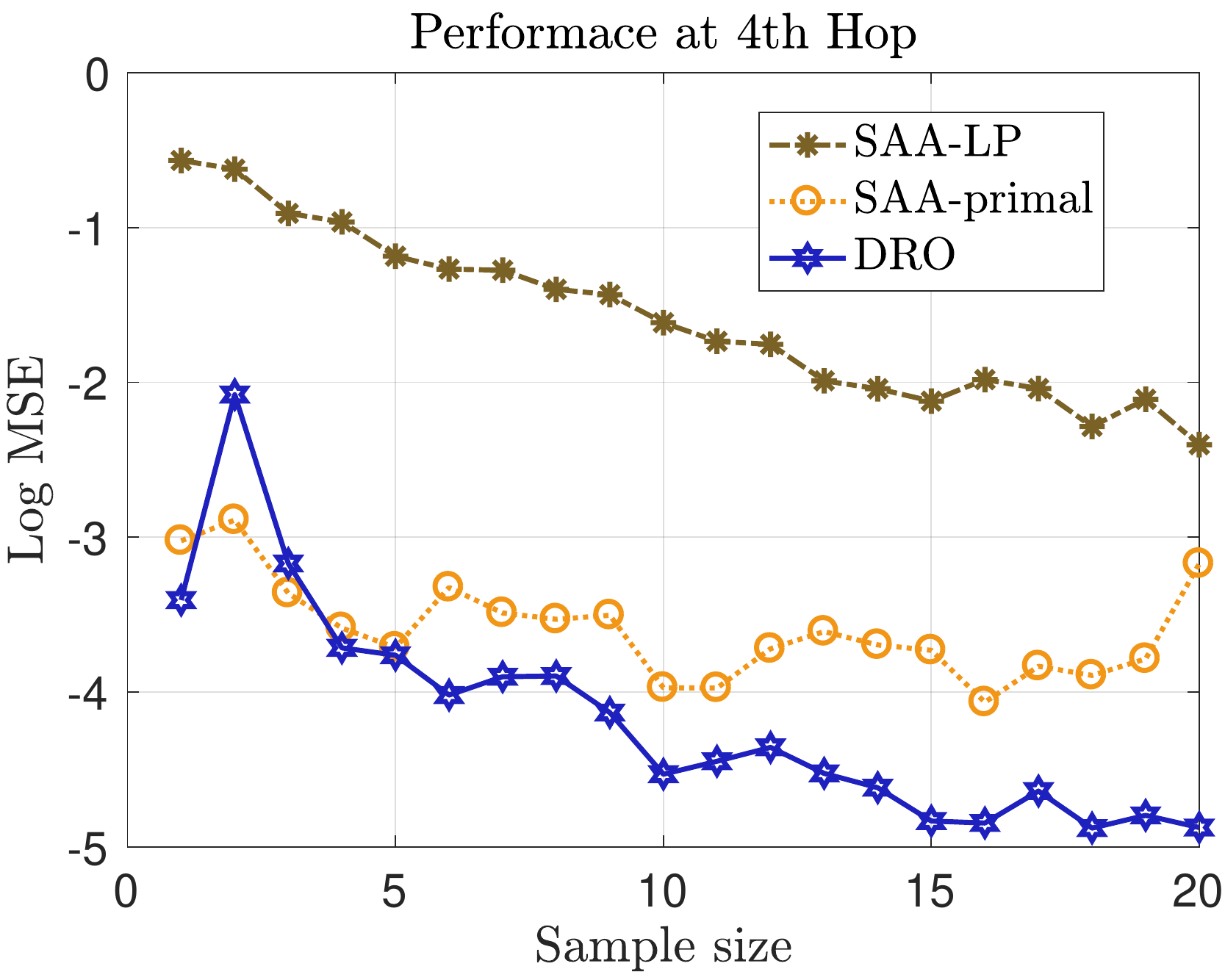}
\includegraphics[width=0.23\textwidth]{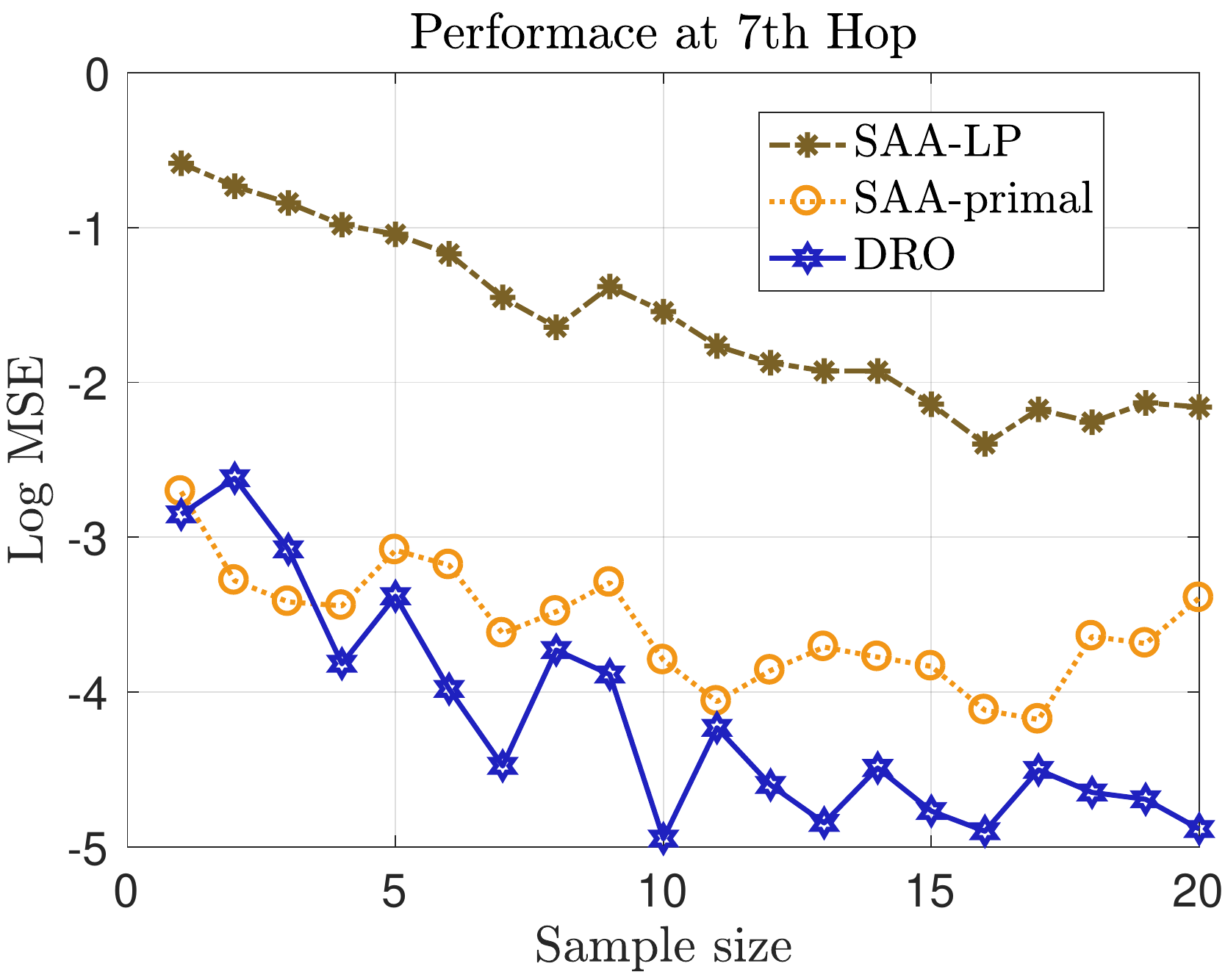}
\includegraphics[width=0.23\textwidth]{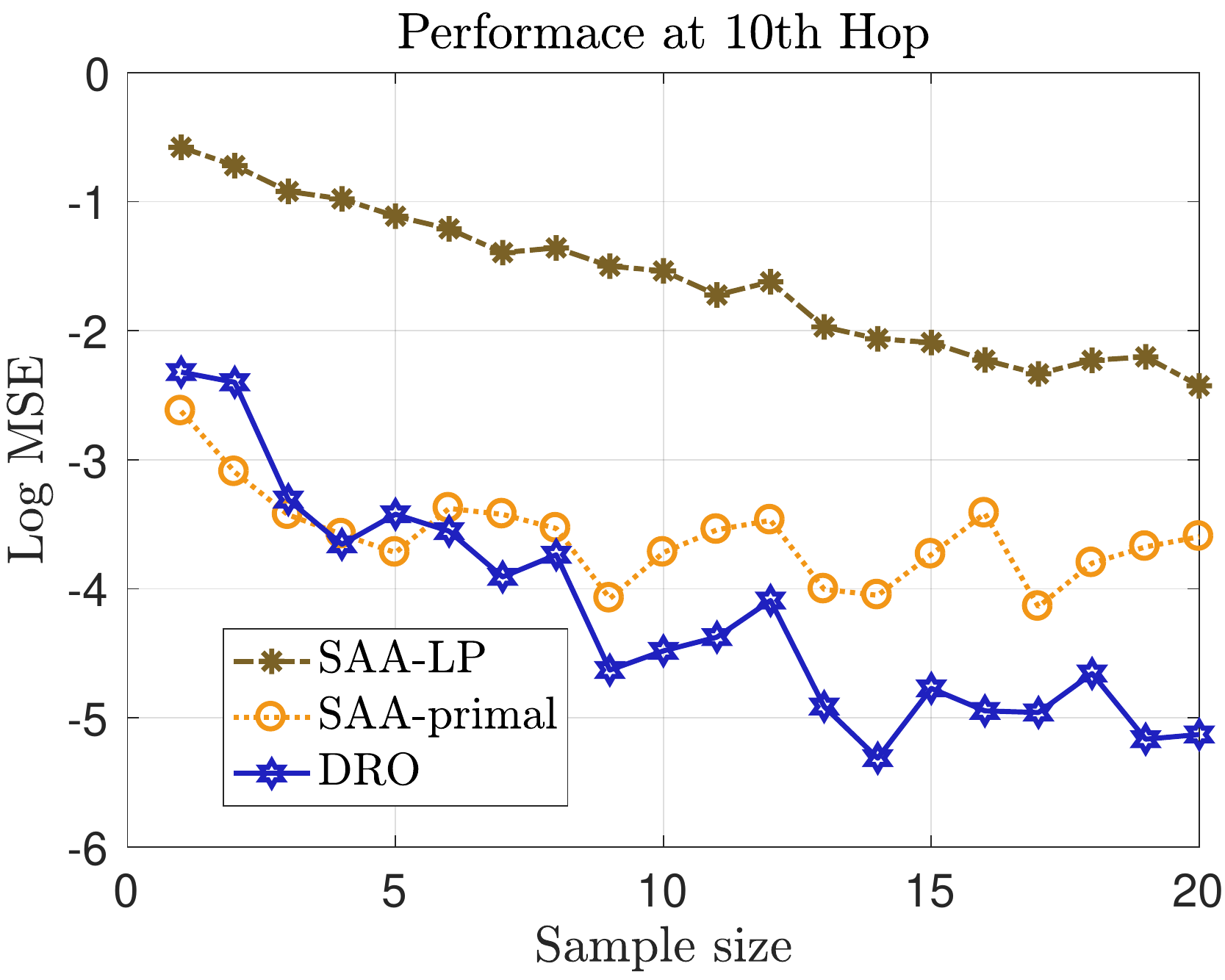}
\caption{Plots for the performance of three different methods with respect to the sample size.
For each fixed sample size the experiment is repeated for 10 independent trials.}\label{fig:sample:throughput}
\vskip -1.5em
\end{figure*}
The radius size $\rho_i$ for the ambiguity set $\mathcal{P}_i$ quantifies the discrepancy between the underlying input rank distribution and the empirical distribution for $i=1,2$.
Provided that the radius size $\rho_i$ is chosen judiciously, our designed ambiguity set can be viewed as the confidence set that contains the underlying input rank distribution with high probability.
As a result, the obtained optimal solution in  \eqref{eq:DRO:IP} constitutes the lower confidence bound of the optimal throughput when the input rank distribution is precisely known.
This idea is mathematically formulated as follows.


\begin{proposition}[Asymptotic Out-of-sample Guarantee~{\cite[Theorem~1]{sommerfeld2017inference}}]\label{Proposition:asymptotic:Wasserstein}
Define a convex set
\[
\Phi^*=\{
u\in\mathbb{R}^{M+1} \colon
u_{r}-u_{r'}\le |r-r'|,\ \forall r,r'\in[M]
\}
\]
and a multi-nomial covariance matrix $\Sigma(\mathfrak{h})\in\mathbb{R}^{(M+1)\times(M+1)}$ by
\[
\Sigma(\mathfrak{h})_{r,r'}=
	\begin{cases}
h_{r}(1-h_r)&\text{if }r=r';\\
-h_rh_{r'}&\text{if }r\ne r'.
	\end{cases}
\]
Define the Gaussian variable $G\sim\mathcal{N}(0,\Sigma(\mathfrak{h}))$.
Then with the sample size $N\to\infty$, we have the weak convergence
\[
	N^{1/2}W(\h, \hath_N)
\implies
\max_{u\in\Phi^*}G\trans u.
\]
\end{proposition}

Based on the asymptotic convergence in Proposition~\ref{Proposition:asymptotic:Wasserstein}, we can choose the radius size $\rho_i=\rho, i=1,2$ such that
\begin{align*}
\Pr\left(
W(\h, \hath_N)>\rho
\right)&=
\Pr\left(
N^{1/2}W(\h, \hath_N)>N^{1/2}\rho
\right)\\
&\approx
\Pr
\left(
X>N^{1/2}\rho
\right)\le 1-\eta,
\end{align*}
where $X\sim \max_{u\in\Phi^*}G\trans u$ denotes the limiting distribution.
In other words, if $N^{1/2}\rho$ is chosen to be the $(1-\eta)$-quantile of the limiting distribution, then asymptotically the optimal value in \eqref{eq:DRO:IP} constitutes as a $(1-\eta)$-confidence lower bound of the optimal throughput.
The distribution of $X$ involves the information about the underlying input rank distribution, which cannot be obtained precisely.
In order to choose the radius size in practical experiments, we approximate the distribution $X$ with $\hat{X}_N$ by replacing the covariance matrix $\Sigma(\mathfrak{h})$ with the empirical estimate $\Sigma(\hath_N)$.
Since the density function for $\hat{X}_N$ is intractable, we approximate the probability by generating $L$ i.i.d. samples $\{x_{\ell}\}_{\ell=1}^L$ from $\hat{X}_N$ and taking $N^{1/2}\rho$ to be the empirical $(1-\eta)$-quantile of $\{x_{\ell}\}_{\ell=1}^L$.
It is also of research interest to explore the finite-guarantees of the ambiguity sets. However, we observe that the choice of the radius is too conservative compared to the asymptotic choice of radius size in Proposition~\ref{Proposition:asymptotic:Wasserstein}.

%
%
%

\section{Experiment Results}

Now we evaluate the performance of DRO for the task of adaptive recoding
numerically.
We set $t_{\text{avg}}=16$, $M=16$, and $\eta=0.95$ throughout this section unless otherwise specified.
Throughout the simulation, we assume that adaptive recoding is deployed for independent packet loss channels with loss rate $0.2$, so the expected rank function $E_r(\cdot)$ can be formulated with a binomial distribution as discussed in \cite{adaptive,uni}.
Given an optimized recoding policy, define its \emph{effective throughput} as 
	$\frac{\mathbb{E}_{r\sim\mathfrak{h}}[E_r(t_r)]}{M}
	\min\left(
	1, \frac{t_{\text{avg}}}{\mathbb{E}_{r\sim\mathfrak{h}}~[t_r]}
	\right)$,
which quantifies the out-of-sample performance for solving \eqref{eq:IP}.
For benchmark comparison, we also study the effective throughput for the SAA method, which directly solves \eqref{eq:IP} while replacing $\mathfrak{h}$ with the estimated empirical distribution $\hath_N$.
We use two approaches to solve this problem.
The first one is by the primal solver outlined in~\cite{uni}, called the \emph{SAA-primal} approach.
The second one is to solve the LP reformulation of SAA directly, called the \emph{SAA-LP} approach, which can be implemented more efficiently.

Fig.~\ref{fig:sample:throughput} reports the performance of recoding policies under different methods across different numbers of collected samples, where the four subfigures correspond to the $1$-st, the $4$-th, the $7$-th, and the $10$-th communication links within a line network. 
Policies are evaluated based on the logarithm of the mean squared error~(MSE) between their effective throughput and the optimal throughput.
We can see that the DRO method outperforms other methods in terms of the MSE criterion. 
Moreover, the MSE for the SAA-primal method is smaller than that of the SAA-LP method, which also justifies that the tuning procedure \cite{uni} applied in the SAA-primal approach makes the obtained recoding policy more robust.

\begin{figure}
\includegraphics[width=0.23\textwidth]{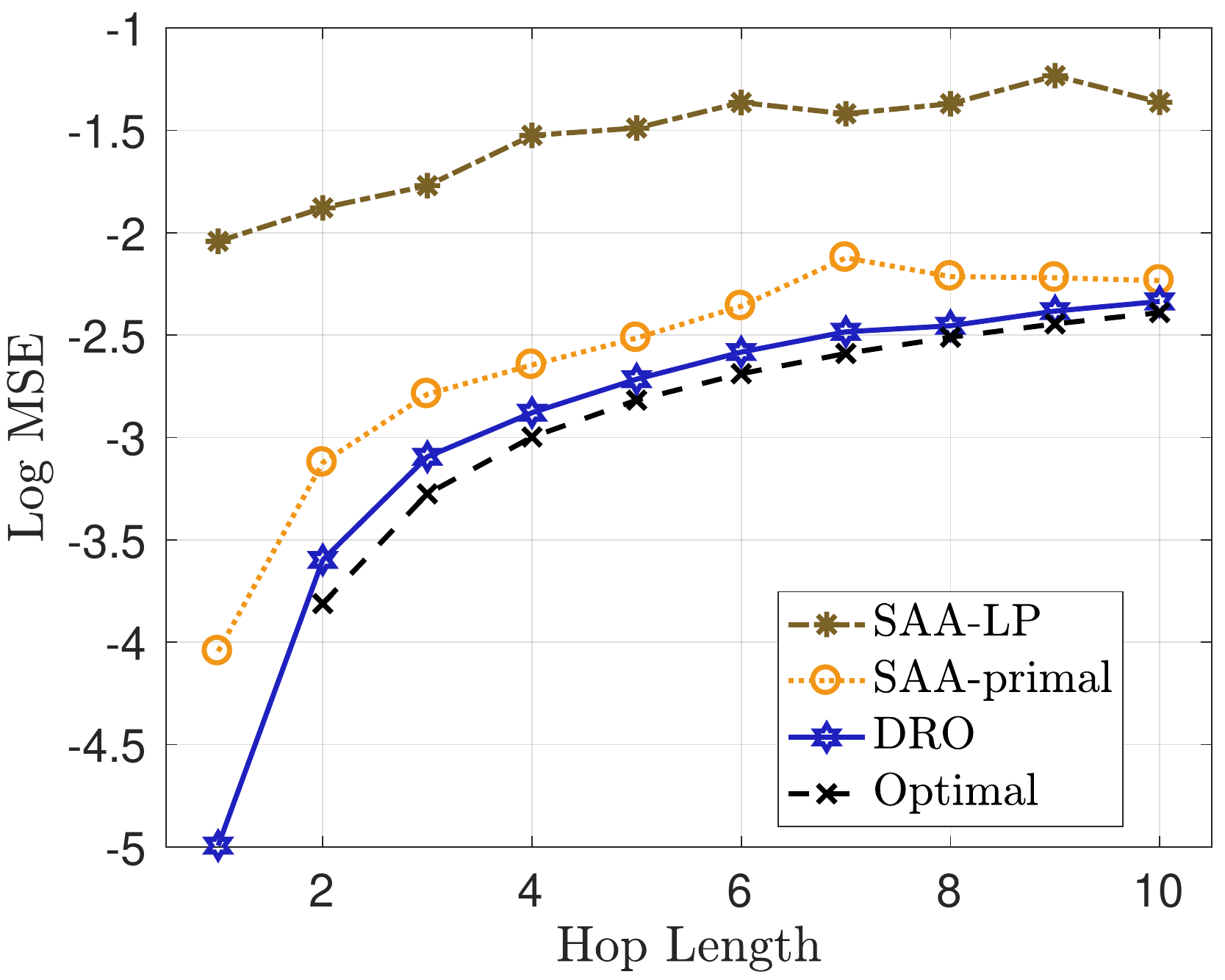}
\includegraphics[width=0.23\textwidth]{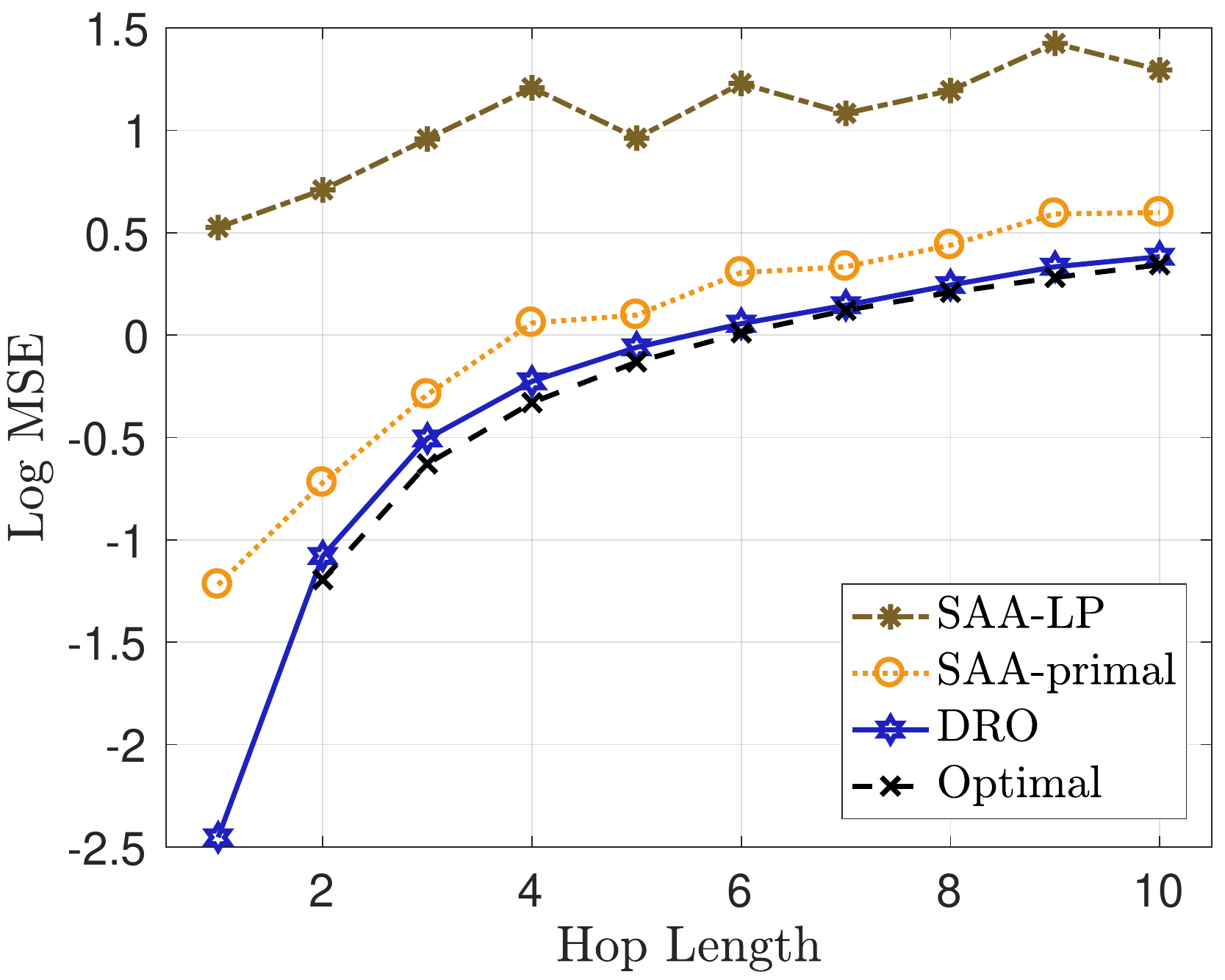}
\caption{Plots for the performance of three different methods under two different line networks with ten hops.
Fig. a) corresponds to the case where $t_{\text{avg}}=16$, and Fig. b) corresponds to the case where $t_{\text{avg}}=20$.}\label{fig:sample:throughput:hop}
\vskip -1.5em
\end{figure}

Fig.~\ref{fig:sample:throughput:hop} reports the performance of recoding policies from different methods under two different line networks with ten hops.
We choose $t_{\text{avg}}=16$ for the first line network and $t_{\text{avg}}=20$ for the second one.
The number of received packets at the intermediate nodes is set to be $N=15$.
The evaluation criterion is chosen to be the logarithm of MSE between the optimal throughput at the first hop and the estimated throughput at the other hops.
We can see that compared to the optimal throughput which the rank distribution is precisely known, the sub-optimality gap for the DRO method is the smallest.

\section{Concluding Remarks}

This paper proposed a distributionally robust optimization framework for adaptive recoding provided that the input rank distribution is not precisely known.
	Generalizing this approach into other settings, such as adaptive recoding under imperfect information about the communication channels, is of research interest.
	In the future, we will also study the optimal selection of the size of the ambiguity sets leveraging tools from non-asymptotic statistics.

\ifnum\paperversion=2

\appendix

\section{Proof of Technical Results}\label{Appendix:proof}

\begin{IEEEproof}[Proof of Theorem~\ref{Theorem:DRP:reformulate:1}]
Since the constraint for Problem~\eqref{eq:DRO:IP} is independent of the decision variable $\mathfrak{h}$, this problem can be reformulated as
\begin{subequations}
\begin{align}
&\sup_{\bm t\in\mathcal{T}}~\mathcal{U}(\bm t\mid\mathcal{P}_1)\label{Eq:DRO:simple:1}\\ 
&\mathcal{T}=\bigg\{
\{t_r\}_{r\in[M]}:~t_r\ge0, 
\mathcal{E}(\bm t\mid\mathcal{P}_2)\le t_{\text{avg}}
\bigg\}.\label{Eq:DRO:simple:2}
\end{align}
\end{subequations}
By the duality result in \cite{gao2016distributionally}, the worst-case expected value of recoded packets has the equivalent formulation:
\begin{multline*}
\mathcal{E}(\bm t\mid\mathcal{P}_2)\\
=
\inf_{\lambda_{0,2}\ge0}~\left\{\lambda_{0,2}\rho_2 + \frac{1}{N}\sum_{j=1}^N\sup_{r\in[M]}\bigg(
t_r - \lambda_{0,2}|r - \hat{r}_j|
\bigg)\right\}.
\end{multline*}
Hence, the constraint set $\mathcal{T}$ can be reformulated as
\begin{align*}
\mathcal{T}&=
\bigg\{
\{t_r\}_{r\in[M]}:~t_r\ge0, 
\exists \lambda_{0,2}\ge0\text{ such that }\\
&\qquad
\lambda_{0,2}\rho_2 + \frac{1}{N}\sum_{j=1}^N\sup_{r\in[M]}\bigg(
t_r - \lambda_{0,2}|r - \hat{r}_j|\bigg)\le t_{\text{avg}}
\bigg\}.
\end{align*}
Similarly, the objective function $\mathcal{U}(\bm t\mid\mathcal{P}_1)$ can be expressed as
\begin{multline*}
\mathcal{U}(\bm t\mid\mathcal{P}_1)\\
=
\sup_{\lambda_{0,1}\ge0}\bigg\{
-\lambda_{0,1}\rho_1 + \frac{1}{N}\sum_{j=1}^N\inf_{r\in[M]}\bigg(
E_r(t_r) + \lambda_1 |r - \hat{r}_j|
\bigg)
\bigg\}.
\end{multline*}
Combining the reformulations of $\mathcal{T}$ and $\mathcal{U}(\bm t\mid\mathcal{P}_1)$ completes the proof.
\end{IEEEproof}
\begin{IEEEproof}[Proof of Theorem~\ref{Theorem:E_r:piecewise:linear}]
	For non-negative integers $i$, define $f_t(i) := E_r(i) + (t-i) \Delta_{r,i}$.
	Refer to the definition of $E_r(t)$, we can see that $E_r(t) = f_t(i)$ when $t \in [i, i+1)$, i.e., $i = \lfloor t \rfloor$.
	In this case, we can reorder the terms to see that $E_r(t) = \Delta_{r,i} t + \zeta_{r,i}$.

	To show that $f_t(i)$ achieves the minimum when $i = \lfloor t \rfloor$, we consider the cases when $t \notin [i, i+1)$.
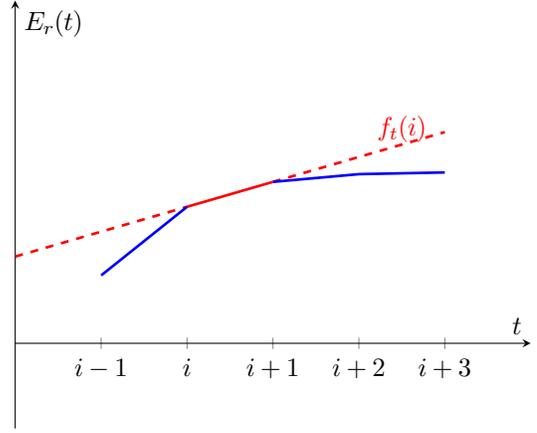
\begin{figure}[H]
\centering
\begin{tikzpicture}[scale=1]
\begin{axis}[my style, xtick={1,2,3,4,5}, ytick={0},
xticklabels={$i-1$, $i$, $i+1$, $i+2$, $i+3$},
xmin=0, xmax=6, ymin=0, ymax=3,
xlabel={$t$}, ylabel={$E_r(t)$}]
\addplot[domain=2:3, red,line width=1pt] {0.29*x+1.01};
\addplot[domain=0:2,dashed, red, line width=1pt] {0.29*x+1.01};
\addplot[domain=3:5,dashed, red, line width=1pt] {0.29*x+1.01};
\addplot[domain=1:2, blue, line width=1pt] {0.8*x-0.01};
\addplot[domain=3:4, blue, line width=1pt] {0.09*x+1.61};
\addplot[domain=4:5, blue, line width=1pt] {0.02*x+1.89};
\node[] at (axis cs: 4.5,2.5) {$\color{red}f_t(i)$};
\end{axis}
\end{tikzpicture}
\caption{Plot of the function $E_r(t)$ versus $t$. The red line represents the plot for $f_t(i)$ over $t$, and the solid line represents the plot for $E_r(t)$ over $t$. As we can see, $E_r(t)=f_t(i)$ if $t\in[i,i+1)$ and otherwise $E_r(t)\le f_t(i)$. Hence we conclude that $E_r(t)=\min_{i\in[i_{\max}^r]}f_t(i)$.}
\end{figure}

	Case I: $t < i$.
	We have $\Delta_{r,i} \le \Delta_{r, i-1} \le \ldots \le \Delta_{r,\lfloor t \rfloor}$.
	Then,
	\begin{IEEEeqnarray*}{Cl}
		& f_t(i)\\
		= & E_r(i) + (t-i) \Delta_{r,i}\\
		= & E_r(t) + (1-(t-\lfloor t \rfloor)) \Delta_{r,\lfloor t \rfloor} + \sum_{k = \lfloor t \rfloor+1}^{i-1} \Delta_{r,k} + (t-i) \Delta_{r,i}\\
		\ge & E_r(t) + (1-(t-\lfloor t \rfloor)) \Delta_{r,i} + \sum_{k = \lfloor t \rfloor+1}^{i-1} \Delta_{r,i} + (t-i) \Delta_{r,i}\\
		= & E_r(t).
	\end{IEEEeqnarray*}

	Case II: $t \ge i+1$.
	We have $\Delta_{r,i} \ge \Delta_{r,i+1} \ge \ldots \ge \Delta_{r,\lfloor t \rfloor}$.
	Then,
	\begin{IEEEeqnarray*}{Cl}
		& f_t(i)\\
		= & E_r(i) + (t-i) \Delta_{r,i}\\
		= & E_r(t) - \sum_{k = i}^{\lfloor t \rfloor-1} \Delta_{r,k} - (t-\lfloor t \rfloor) \Delta_{r,\lfloor t \rfloor} + (t-i) \Delta_{r,i}\\
		\ge & E_r(t) - \sum_{k = i}^{\lfloor t \rfloor-1} \Delta_{r,i} - (t-\lfloor t \rfloor) \Delta_{r,i} + (t-i) \Delta_{r,i}\\
		= & E_r(t).
	\end{IEEEeqnarray*}

	The above two cases show that $f_t(i) \ge E_r(t)$ for all $i \neq \lfloor t \rfloor$.
	Thus, the proof is done.
\end{IEEEproof}

\begin{IEEEproof}[Proof of Corollary~\ref{Corollary:DRO:LP:reformula}]
By introducing epi-graphical slack variables $\{\lambda_{j,1}\}_{j=1}^N$ and $\{\lambda_{j,2}\}_{j=1}^N$ into the formulation in Theorem~\ref{Theorem:DRP:reformulate:1}, it suffices to consider the following problem:
\begin{subequations}
\begin{align*}
\max_{\substack{
\bm t\ge0\\
\lambda_{0,1}\ge0, \lambda_{1,1},\ldots,\lambda_{N,1}\\
\lambda_{0,2}\ge0, \lambda_{1,2},\ldots,\lambda_{N,2}\\
}}&\quad -\lambda_{0,1}\rho_1 + \frac{1}{N}\sum_{j=1}^N\lambda_{j,1}\\
\mbox{s.t.}&\quad \lambda_{0,2}\rho_2 + \frac{1}{N}\sum_{j=1}^N\lambda_{j,2}\le t_{\text{avg}}\\
&\quad \lambda_{j,1}\le E_r(t_r) + \lambda_{0,1} |r - \hat{r}_j|,\\
&\qquad\qquad\ j=1,\ldots,N, r\in[M]\\
&\quad \lambda_{j,2}\ge t_r - \lambda_{0,2} |r - \hat{r}_j|,\\
&\qquad\qquad\ j=1,\ldots,N, r\in[M]
\end{align*}
\end{subequations}
Substituting $E_r(t)$ with the piece-wise linear function stated in Theorem~\ref{Theorem:E_r:piecewise:linear} gives the desired result.
\end{IEEEproof}

\begin{IEEEproof}[Proof of Proposition~\ref{Proposition:regularization}]
By applying the result in \cite[Proposition~6]{wang2021reliable},
as long as $\rho_1<\bar{\rho}_1$, the objective function in \eqref{eq:DRO:IP} can be reformulated as
\[
\mathcal{U}(\bm t\mid\mathcal{P}_1)
=
\mathbb{E}_{r\sim\widehat{\mathfrak{h}}}~[E_r(t_r)] - \rho_1\|E^{\bm t}\|_{\text{Lip}, \widehat{\mathfrak{h}}}.
\]
Similarly, as long as $\rho_2<\bar{\rho}_2$, the left-hand side in the constraint becomes
\[
\mathcal{E}(\bm t\mid\mathcal{P}_2)
=
\mathbb{E}_{r\sim\widehat{\mathfrak{h}}}~[t_r] + \rho_2\|\bm t\|_{\text{Lip}, \widehat{\mathfrak{h}}}.
\]
The proof is completed.
\end{IEEEproof}

\fi

\clearpage
\balance
\bibliographystyle{IEEEtran}
\bibliography{DRO-bib}

\end{document}